\documentclass[journal]{IEEEtran}
\usepackage{cite}

\ifCLASSINFOpdf
   \usepackage[pdftex]{graphicx}
\else
\fi
\usepackage{amsmath}
\usepackage{amssymb}
\ifCLASSOPTIONcompsoc
  \usepackage[caption=false,font=normalsize,labelfont=sf,textfont=sf]{subfig}
\else
  \usepackage[caption=false,font=footnotesize]{subfig}
\fi

\usepackage{color}
\usepackage[dvipsnames]{xcolor}

\begin{document}
%

\title{Remote-Strong-Grid-Point-Based Synchronization Strategy with Fault Ride-Through Capability for Distributed Energy Resources Connected to Weak Grids}

%
%

\author{Runfan Zhang
        and~Branislav Hredzak,~\IEEEmembership{Senior Member,~IEEE}
\thanks{Runfan Zhang and Branislav Hredzak are with the School of Electrical Engineering and Telecommunications, University of New South Wales, Sydney, NSW 2052 Australia (email: runfan.zhang@student.unsw.edu.au; b.hredzak@unsw.edu.au;).}}
\maketitle

\begin{abstract}
This paper proposes a novel strategy for the current injection based control for distributed energy resources connected to weak grids through a voltage source converter experiencing faults. The current injection controller is no longer synchronized with the point of common coupling at which the measured voltage signal can be severely affected by faults in the weak grid, but with the strong grid point at which the voltage is rigid. It is shown that the  phase difference between the voltage source converter and the strong grid voltages caused by the long power lines does not affect the power control. Furthermore, a time delay compensation method which tolerates communication time delay introduced by the transmission of the synchronization signal from the strong grid point is proposed. The performance of the proposed control strategy is verified with an RTDS Technologies real-time digital simulator using switching converter models.
\end{abstract}


%
\IEEEpeerreviewmaketitle

\section{Introduction}
%
%
%
%
\IEEEPARstart{T}{he} weak networks are characterized by short circuit ratio (SCR) below 1 p.u. and X/R ratios at around 1 (rather than $>5$) \cite{SCR,weak_grid}. These characteristics present technical issues that are becoming a significant drag on further integration of renewable energy sources. The problem is compounded by the poor performance of existing inverter connected generation in the network that has also contributed to current and future challenges.

To interface the renewable energy sources to the main grid, the current injection control (grid feed control) tracks the real and reactive power set-points and synchronizes to the point of common coupling. The current-controlled mode requires phase lock loop (PLL) synchronization with the AC voltage at the connection point in order to control the active and reactive powers exchanged with the grid. Under weak grid, many conditions such as converter start, current reference change and grid faults will cause severe interference to PLL. As a result, the current-controlled mode can suffer from instability issues – improvements can be made by reduction of the bandwidth of the PLL and acceleration of the voltage regulator response \cite{PE}. When the voltage source converter (VSC) connecting the renewable generation source to the weak grid is synchronized  with the point of common coupling (PCC) the control system can fail to maintain stable operation of the VSC when there is a fault in the weak grid.

Specifically, the weak grid can cause the following problems in the current-mode controlled VSC \cite{wk_issue1,wk_issue2}:
\begin{enumerate}
	\item Under the weak grid conditions, the voltage fluctuation at the PCC will disturb the PLL leading to a transient response. As a result, the system cannot achieve desired control performance.
	\item Current control interacts with the PLL via the voltage at the PCC. The PLL dynamics may deteriorate the grid current control and even result in the system instability.
	\item A PLL with a high bandwidth improves the dynamic response of the grid-connected converters; nevertheless, it may result in interactions between the converter and grid and increase the negative real part of the converter output impedance, which will destabilize the system under the weak grid conditions.
	\item By slowing down the PLL, the disturbance of the grid current control during the transient process can be effectively suppressed. However, a slow PLL increases the response time of the grid current and hence convergence to a steady state, which is not beneficial for the fault ride-through.
	\item Consequently, it is not practical to re-tune the PLL parameters to enhance the grid current dynamics, because it is limited by the PLL performance trade-offs.
	\item As a consequence, some literature proposes additional control strategies to alleviate the negative impacts of PLL: i) a supplementary controller is integrated with the outer control loops \cite{wk_issue3}, ii) an impedance conditioning term is introduced in the PLL input voltage \cite{wk_issue4}, iii) an impedance-phased compensation control strategy is proposed \cite{wk_issue5}.

\end{enumerate}
\begin{figure*}[!t]
	\centering
	\includegraphics[width=6in]{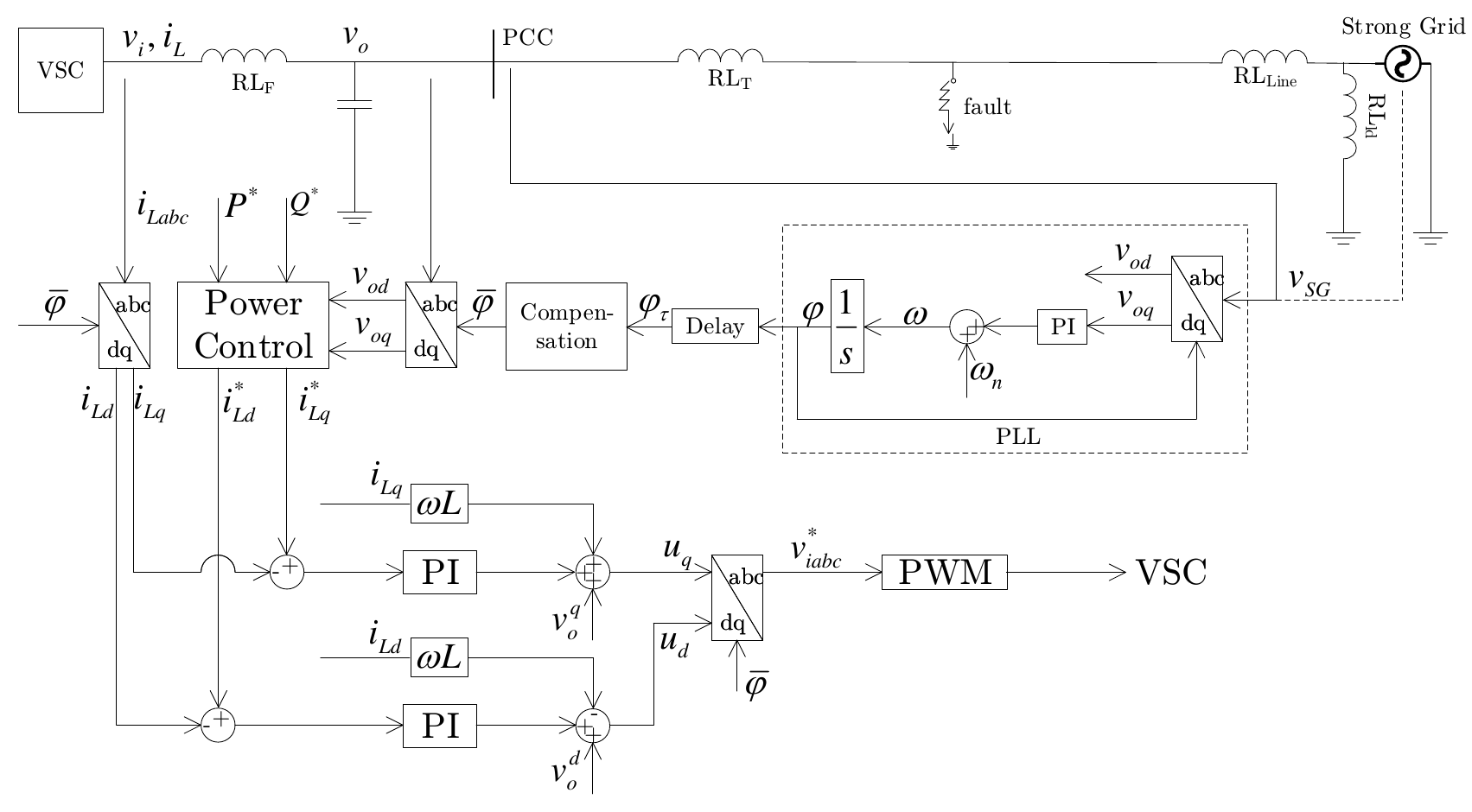}
	\caption{Illustration of a weak grid and the proposed control strategy. The VSC is interfaced to the PCC via an LC filter. A transformer connects long-distance power line (the weak grid) to the strong grid. The PLL is synchronized to the strong grid voltage $v_{SG}$ (or the PCC ). $\varphi$ can be delayed due to communication.} 
	\label{fig_wg}
\end{figure*}
Thus, with the trend of increasing penetration of renewable energy sources (such as solar generation, wind turbines) and energy storage systems \cite{wk_rnwsrc}, the stability of the weak grid connected VSCs is of paramount importance for stable and reliable operation of next generation power systems. 

Motivated by the above discussion, this paper proposes a novel strategy which synchronizes the VSC's PLL to the strong grid point voltage rather than to the PCC voltage. Furthermore, the synchronization to the strong grid point voltage does not affect the local VSC power control. Also, the communication time delay introduced by the transmission of the the strong grid point voltage data can be compensated. The salient features of this paper are:
\begin{enumerate}
	\item The PLL of the local VSC control is synchronized to the strong grid point voltage and achieves the stable operation and improvement in the output voltage, and real and reactive powers dynamics;
	\item The strong grid synchronization control strategy improves the VSC outputs dynamics and recovery after faults;
	\item The strong grid point synchronization does not affect the local powers control;
	\item The proposed time delay compensation method eliminates effect of the time delays on the proposed control strategy.
\end{enumerate}

The paper is organized as follows. Section II describes the weak grid and proposed control method. Real-time simulations verifying the proposed strategy are presented in Section III. Section IV summarizes the paper. 

\section{Weak Grid and Current Injection Control}\label{}
\subsection{Weak Grid}
For stable operation, VSCs are used to interface renewable energy sources to grid. However, the renewable generation sources are commonly located in rural, isolated areas and are connected to the main grid (strong grid) through a long distance power line with high impedance. This type of grid is called a weak grid. Without a proper control strategy, the high impedance of the weak grid increases the transient response after a fault in the grid, and can even result in instability and collapse of the voltage and real power and reactive power injection.

The weak grid is illustrated in Fig. \ref{fig_wg}. The VSC is interfaced to the PCC via an LC filter. A transformer connects long-distance power line (weak grid) to the strong grid. The long-distance power line's impedance can be very high \cite{WK_PLL}.

\subsection{Current Injection Control}
The current injection control is using a PLL to synchronize with the PCC and inject power to the grid. Conventionally, the PLL generates the phase angle $\varphi$ based on the voltage signals at the PCC. Based on the $\varphi$, the measured three-phase inductor currents $i_{Labc}$ and output voltages $v_{oabc}$ are transformed into the \textit{d-q} frame. Then, based on the \textit{d-q} frame voltages, and real and reactive powers set points, the inductor current references in \textit{d-q} frame are calculated. The powers set points can be determined by the power network demands for the energy storage and the maxim power point tracking for renewable energy sources. Alternatively, an output voltage restoration control can be introduced to set the reactive power reference in order to recover the output voltage by the reactive power injection. Subsequently, PI controllers regulate the inductor currents to inject the power into grid and generate the VSC voltage references. Feed forward voltage terms are included in the current control loops to improve the robustness and accelerate the current control responses. The voltage references are then sent to the PWM generator.

In Fig. \ref{fig_wg}, the local control system measures the three phase output voltages and inductor currents $v_{oabc}$ and $i_{Labc}$. Then, the power invariant \textit{d-q} transformation is applied to obtain the output voltages and inductor currents in the \textit{d-q} frame, $v_{odq}$ and $i_{Ldq}$, based on $\varphi$ generated by PLL synchronized with the voltage signals at the PCC. Assume that the real power and reactive power set points are $P^*$ and $Q^*$ respectively. The power controller calculates the inductor current references as follows,
\begin{equation}\label{eq_pwrc}
\begin{array}{l}
i_{Ld}^* = \frac{{{v_{od}}{P^*} - {v_{oq}}{Q^*}}}{{v_{od}^2 + v_{oq}^2}},\\
i_{Lq}^* = \frac{{{v_{oq}}{P^*} + {v_{od}}{Q^*}}}{{v_{od}^2 + v_{oq}^2}}.
\end{array}
\end{equation}
The \textit{d-q} inductor currents are controlled by PI controllers and set the VSC voltage reference $u_{dq}$ in the \textit{d-q} frame. The \textit{dq-abc} transformation transforms the \textit{d-q} voltage references to the \textit{abc} frame for the PWM generator.

Commonly, the PCC voltage based PLL output $\varphi$ is able to synchronize the output voltage with the grid with a smaller line impedance. However, to transmit the power generated by the renewable energy sources to the main grid, the power lines are long due to the fact that these energy sources are located in the rural areas. As a result, the output voltage and powers can be affected by the faults in the weak grid. The higher the line impedance the more obvious the effect, and ultimately the VSC cannot synchronize with the grid.

\subsection{Proposed Current Injection Control}
To stabilize the VSC output voltage, and the real and reactive powers injection after faults in the weak grid, it is proposed that the PLL is no longer synchronized to the PCC voltage, but the strong grid point voltage which results in robust fault ride-through capability. As shown in Fig. \ref{fig_wg}, the PLL obtains $\varphi$ from the strong grid voltage $v_{SG}$. However, since the renewable energy source are located in rural areas, the communication between the local controller and the strong grid can be delayed. A simple method can be used to compensate and remove the delay effects.

\subsection{Delay Compensation}\label{sec_TDC}
In Fig. \ref{fig_wg}, assume that  $\varphi_\tau$ is the delayed $\varphi$. This corresponds to a phase shift with respect to the strong point voltage. Since $0 \le {\varphi _\tau } < 2\pi $, the time delay is within the range $0 \le \tau  < \frac{1}{f_N}$, where $f_N$ is the rated frequency of the power system. The longest, worst case value of the time-delay is $\tau  = \frac{1}{{2f_N}}$. To compensate the time delay, it is assumed that the transmitted data is time stamped and hence the time delay is known by the local controller. The local controller calculates the compensated angle as $\bar \varphi  = {\varphi _\tau } + \Delta \varphi  = {\varphi _\tau } + \tau \omega_N $, where $\omega_N=2\pi f_N$. $\bar \varphi$ is then used by the \textit{d-q} transformation and to generate the \textit{abc} frame voltage references for the PWM generator.

\subsection{Power Invariant Analysis }
\begin{figure}[!t]
	\centering
	\includegraphics[width=2in]{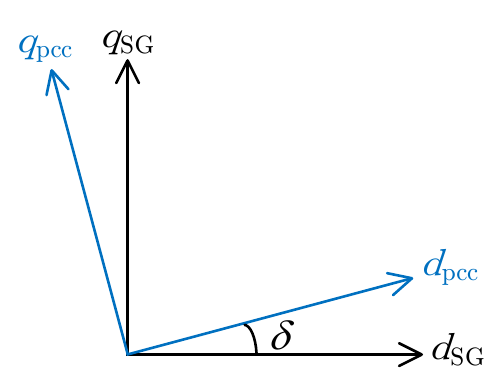}
	\caption{Reference frames at the PCC and strong grid point.}
	\label{fig_pwrinv}
\end{figure}
The application of different PLL angles will not result in the real and reactive power offsets by the power control \eqref{eq_pwrc} and current control system in Fig. \ref{fig_wg}. Denote the PCC voltages and inductor currents in the PCC reference frame ($v_{PCC}$ is used for the PLL synchronization) as ${v_{odq}} = {\left[ {\begin{array}{*{20}{c}}
		{{v_{od}}}&{{v_{oq}}}
		\end{array}} \right]^T}$ and ${i_{Ldq}} = {\left[ {\begin{array}{*{20}{c}}
		{{i_{Ld}}}&{{i_{Lq}}}
		\end{array}} \right]^T}$, and in the strong grid reference frame ($v_{SG}$ used for the PLL synchronization) as $v_{odq}^{SG} = {\left[ {\begin{array}{*{20}{c}}
		{v_{od}^{SG}}&{v_{oq}^{SG}}
		\end{array}} \right]^T}$ and $i_{Ldq}^{SG} = {\left[ {\begin{array}{*{20}{c}}
		{i_{Ld}^{SG}}&{i_{Lq}^{SG}}
		\end{array}} \right]^T}$ respectively.
Based on \cite{ACMG_dq}, the state variables transformation between the two frames can be described as follows,
\begin{equation}\label{eq_Tdq}
	\begin{array}{l}
	f_{dq}^{SG} = T{f_{dq}},\\
	T = \left[ {\begin{array}{*{20}{c}}
		{\cos \delta }&{ - \sin \delta }\\
		{\sin \delta }&{\cos \delta }
		\end{array}} \right],
	\end{array}
\end{equation}
where $\delta$ is the angle of the reference frame of the PCC with respect to the strong grid reference frame as shown in Fig. \ref{fig_pwrinv}; $f_{dq}^{SG}$ and ${f_{dq}}$ can be any state variables based on the strong grid and PCC reference frames respectively. Now, the output powers of VSC can be defined as,
\begin{equation}\label{eq_pwrdq}
	\begin{array}{l}
	P = v_{dq}^T{i_{Ldq}},\;Q = v_{dq}^TM{i_{Ldq}},\\
	M = \left[ {\begin{array}{*{20}{c}}
		0&1\\
		{ - 1}&0
		\end{array}} \right].
	\end{array}
\end{equation}
Then, applying \eqref{eq_Tdq} to the PCC frame based voltages and inductor currents and substituting into \eqref{eq_pwrdq} gives,
\begin{equation}\label{eq_pwrSGdq}
	\begin{array}{l}
	P = v_{dq}^T{i_{Ldq}} = {\left( {{T^{ - 1}}v_{dq}^{SG}} \right)^T}{T^{ - 1}}i_{Ldq}^{SG}\\
	\;\;\; = {\left( {v_{dq}^{SG}} \right)^T}{\left( {{T^{ - 1}}} \right)^T}{T^{ - 1}}i_{Ldq}^{SG}\\
	\;\;\; = {\left( {v_{dq}^{SG}} \right)^T}i_{Ldq}^{SG},\\
	\;Q = v_{dq}^TM{i_{Ldq}} = {\left( {{T^{ - 1}}v_{dq}^{SG}} \right)^T}M{T^{ - 1}}i_{Ldq}^{SG}\\
	\;\;\;\; = {\left( {v_{dq}^{SG}} \right)^T}{\left( {{T^{ - 1}}} \right)^T}M{T^{ - 1}}i_{Ldq}^{SG}\\
	\;\;\;\; = {\left( {v_{dq}^{SG}} \right)^T}Mi_{Ldq}^{SG}.
	\end{array}
\end{equation}
From \eqref{eq_pwrdq} and \eqref{eq_pwrSGdq}, the power calculation is invariant to the different reference frames. Then, using \eqref{eq_pwrc} and the control systems in Fig. \ref{fig_wg}, the strong grid synchronization method can be applied to control the real and reactive power accurately. 

Furthermore, the proposed novel control strategy can also be applied to multi-VSCs by broadcasting the $\varphi$ to all VSCs in the weak grid.

\section{Results}\label{}
\begin{table}[!t]
	\renewcommand{\arraystretch}{1.3}
	\caption{Case Study Parameters-Power System}
	\label{parameters_ps}
	\centering
	\begin{tabular}{l l l l l l}
		\hline
		$V_{BSG}$ &  11kV	&	$V_{BVSC}$	&	3.3kV	& $S_B$	& 5MVA	\\
		$X_L$& 1.442	& $X_C$	&-182.7	&	$X_T$	& 0.05   \\
		$X_{Line}^\alpha$	& 	0.1298&$R_{Line}^\alpha$	& 0.01298&		$X_{Line}^\beta$	& 0.5193\\
		$R_{Line}^\beta$		 	& 	0.05193	&	$R_{Load}$	&	0.25	& $\omega_n$	& 314.2rad/s	\\
		\hline
	\end{tabular}
\end{table}

\textbf{\begin{table}[!t]
		\renewcommand{\arraystretch}{1.3}
		\caption{Case Study Parameters-Control System}
		\label{ctrl_ps}
		\centering
		\begin{tabular}{l l l l l l l l}
			\hline
			$K^P_{PLL}$ &  0.3&$K^I_{PLL}$	&	100	& $K^P_{C}$	& 50&$K^I_{C}$& 2000	\\
			$K^I_{PLL}$& 2000	& $P^*$	&	1	&	$Q^*$	& -0.2   &	$f_{DC/AC}$	& 20kHz   \\
			\hline
		\end{tabular}
	\end{table}
}

The presented real-time simulation results compare performance when the PCC voltage and the strong grid voltage are used for the PLL synchronization and a three-phase line to ground fault is applied.

A 3.3kV VSC interfaces with an 11kV weak grid through a transformer. An LC filter connects the VSC and the transformer. The weak grid has a large line impedance between the transformer and the strong grid. The strong grid has a 0.25 p.u.\footnote{The units in the following parts of the paper are all in p.u..} resistive load. The power set points are $P^*=1$ and $Q^*=-0.2$ in the case studies. Other power system parameters are shown in TABLE \ref{parameters_ps}. Control parameters are shown in TABLE 
\ref{ctrl_ps}. The three-phases line to ground fault exists for 10 cycles and then it is removed. Power line impedances vary in the following case studies. 

Three case studies are presented. In Case A, the line impedance is small and both the PCC and strong grid synchronization can maintain the voltages, and real and reactive powers after the fault. However, in Case B the line impedance is increased and the PCC based synchronization is no longer able to regulate the voltages, and real and reactive powers after the fault. Case C presents simulation results for  the worst-case, longest time delay condition and its compensation.

\subsection{Small Line Impedance}
\begin{figure}[!t]
	\centering
	\includegraphics[width=3.5in]{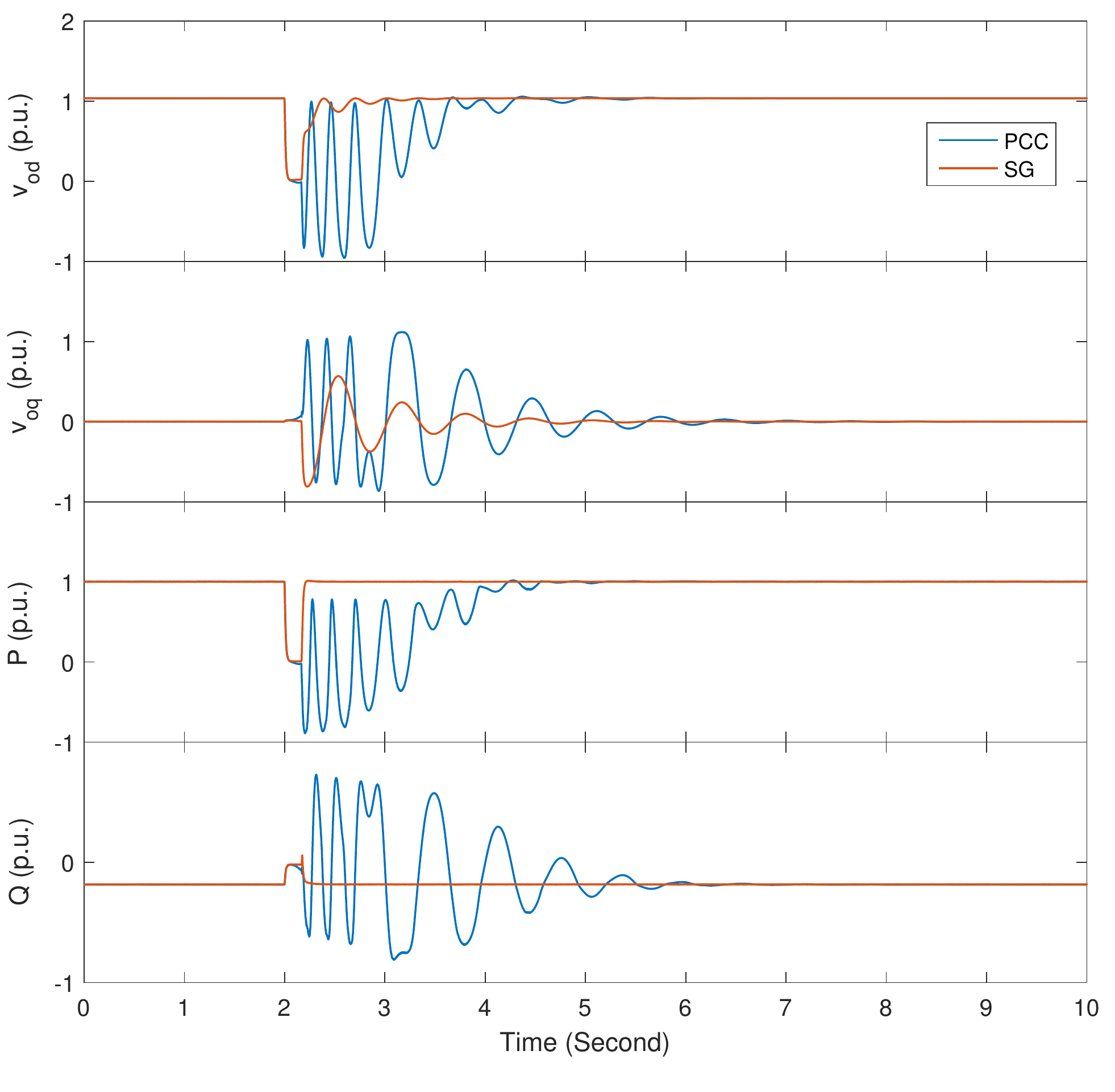}
	\caption{Comparison of voltages, real and reactive powers of the VSC when the PCC voltage and the strong grid voltage are used for the PLL synchronization and a three-phase line to ground fault is applied. Assume no delay, $\tau=0$. The line impedance is small, $R_{Line}^\alpha=0.01298$ and $X_{Line}^\alpha=0.1298$. }
	\label{fig_RX1}
\end{figure}
The power line impedance in this case is  $R_{Line}^\alpha=0.01298$, $X_{Line}^\alpha=0.1298$ and  SCR$=0.5339$. In this case, both synchronization with $v_{PCC}$ and $v_{SG}$ can maintain stable operation. The results are shown in Fig. \ref{fig_RX1}. However, based on Fig. \ref{fig_RX1}, synchronization of PLL to $v_{SG}$ can improve the dynamic performance and shorten the transient time. Furthermore, as described in Section II-E, the power control is reference frame invariant and accurate power control is achieved when synchronizing to $v_{SG}$.

\subsection{Large Line Impedance}
\begin{figure}[!t]
	\centering
	\includegraphics[width=3.5in]{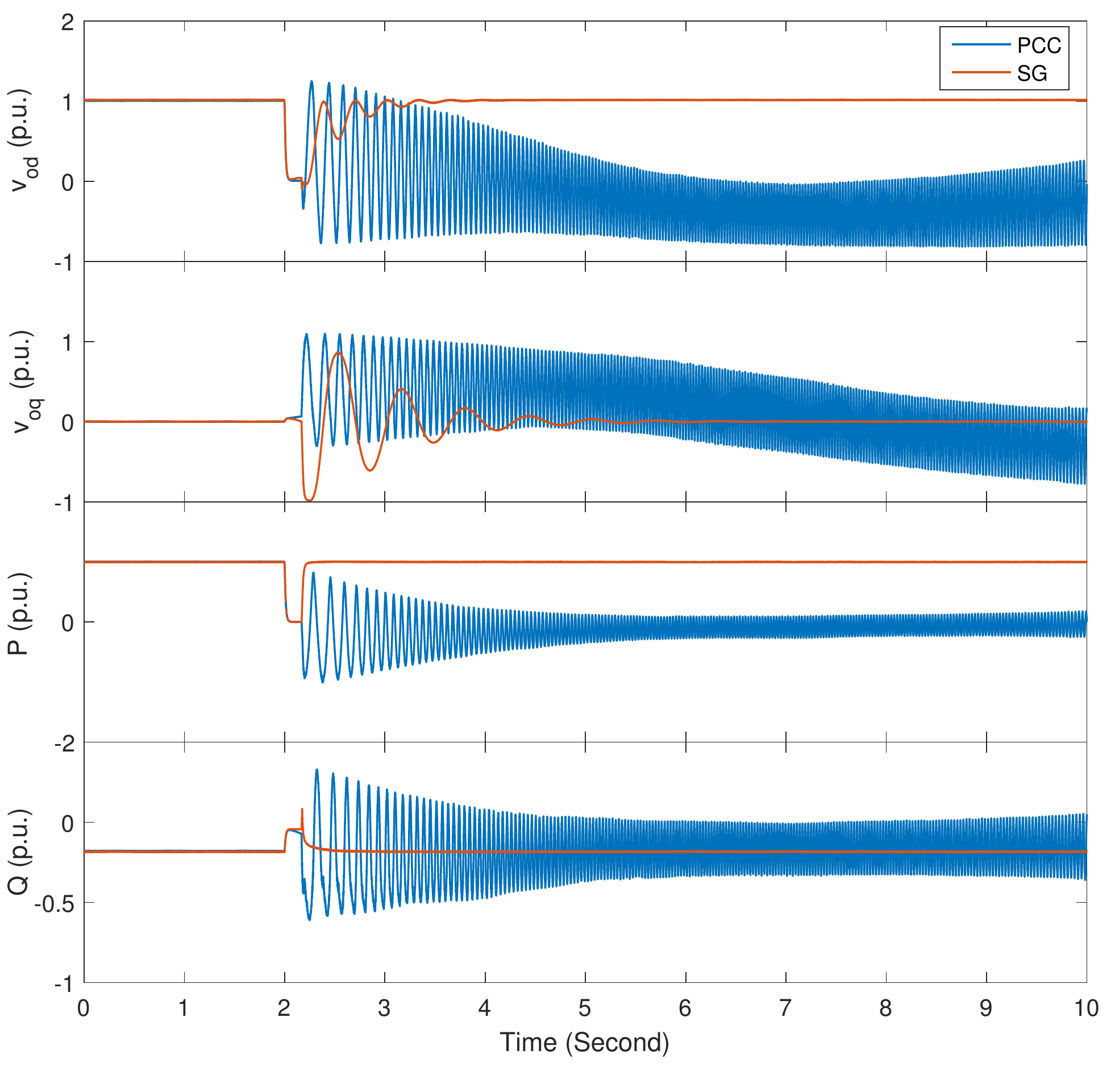}
	\caption{Comparison of voltages, real and reactive powers of the VSC when the PCC voltage and the strong grid voltage are used for the PLL synchronization and a three-phase line to ground fault is applied. Assume no delay, $\tau=0$. The line impedance is large, $R_{Line}^\beta=0.05193$ and $X_{Line}^\beta=0.5193$.}
	\label{fig_RX4}
\end{figure}

Although synchronization of PLL to both $v_{PCC}$ and $v_{SG}$ can be maintained for a small line impedance this is not the case for larger line impedances. In case of a larger line impedance ($R_{Line}^\beta=0.05193$, $X_{Line}^\beta=0.5193$, SCR $=0.4411$) synchronization of PLL to $v_{PCC}$ is no longer stable as shown in Fig. \ref{fig_RX4}. Fig. \ref{fig_RX4} presents the VSC outputs dynamics after the fault when PLL is synchronized to both $v_{PCC}$ and $v_{SG}$. The proposed control strategy provides stable operation after the fault.
\subsection{Time Delay Compensation}
\begin{figure}[!t]
	\centering
	\includegraphics[width=3.5in]{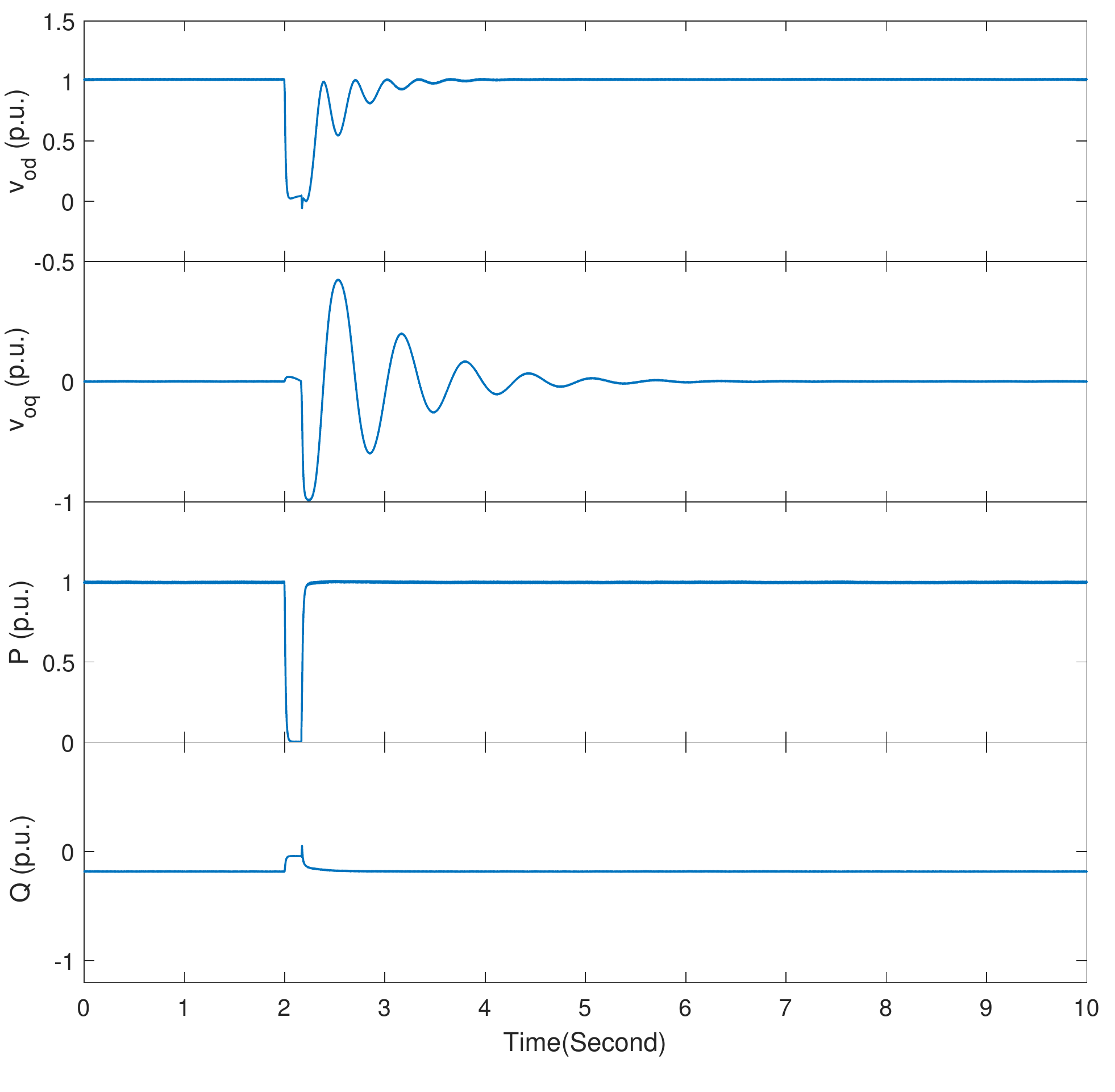}
	\caption{Voltage, real and reactive powers of the VSC when the strong point grid voltage is used for the PLL synchronization and a three-phase line to ground fault is applied. $\varphi$ is time-delayed by  $\tau=0.01$s and compensated as described in Section II-D. The line impedance is large, $R_{Line}^\beta=0.05193$ and $X_{Line}^\beta=0.5193$.}
	\label{fig_RX4TDC}
\end{figure}
In this case study, it is assumed that $\varphi$ is delayed by the longest possible time delay $\tau=0.01$s. Then, the compensation method described  in Section \ref{sec_TDC} is applied to the local control. The results shown in the Fig. \ref{fig_RX4TDC} demonstrate that with the delay compensation the proposed control strategy can maintain stable operation.


\section{Conclusion}\label{Sec_Conclusion}
This paper proposed a novel strategy to stabilize a weak grid connected VSC after a fault by synchronizing the VSC to a strong grid point rather than to the PCC. With the increase of the power transmission line impedance, the proposed control strategy can still maintain stable operation of the VSC after a fault, while the synchronization to the PCC fails. In addition, in the case of a low line impedance, the proposed strong grid point synchronization method  improves the transient dynamics performance after the fault. Furthermore, although different voltage is applied to the PLL, an accurate real and reactive power control is achieved. Real-time simulations on a low voltage weak grid with a switching model of VSC illustrated that the proposed method can maintain stable operation of the VSC after there is a three-phase line to ground fault in the weak grid.


%





\ifCLASSOPTIONcaptionsoff
  \newpage
\fi



\bibliographystyle{IEEEtran}
\bibliography{IEEEabrv,IEEEreference}
%

%











\end{document}